\documentclass[11pt, a4paper, oneside]{article}
\usepackage[section]{placeins}

\usepackage[
  backend=biber,
  style=numeric,
  sorting=none,
  maxbibnames=10,
  maxcitenames=2,
  giveninits=true
]{biblatex}

\DeclareNameAlias{default}{family-given} 
\usepackage{mathtools}

\usepackage{paralist}
\usepackage{setspace}
\usepackage{graphicx}
\usepackage{graphics}
\usepackage{wrapfig}
\usepackage{tabularx}
\usepackage[dvipsnames]{xcolor}
\usepackage[margin=2cm]{geometry}
\usepackage[most]{tcolorbox}
\usepackage{float}
\usepackage[normalem]{ulem}
\definecolor{arsenic}{rgb}{0.23, 0.27, 0.29}
\usepackage[font=small,skip=5pt]{caption}
\usepackage{titlesec}
\usepackage{arydshln}
\usepackage{soul}
\usepackage{subcaption}
\usepackage{adjustbox}
\usepackage{lineno}
\usepackage{multirow}
\usepackage{multicol}
\usepackage{tabularx}
\usepackage{algorithm}
\usepackage{algpseudocode}
\usepackage{listings}
\usepackage{titling}
\usepackage{array}
\usepackage{makecell}
\usepackage{MnSymbol}
\usepackage{scrextend}
\usepackage{caption}
\usepackage{xurl}
\usepackage{hyperref}
\usepackage{siunitx}
\DeclareSIUnit{\Molar}{\textsc{M}}
\DeclareCiteCommand{\citeyearpar}
    {}
    {\mkbibparens{\bibhyperref{\printdate}}}
    {\multicitedelim}
    {}

\usepackage{etoolbox}
\AtBeginEnvironment{tabular}{\small}

\newcounter{systemprompt}
\renewcommand{\thesystemprompt}{\arabic{systemprompt}}


\begin{document}

\newcommand{\RC}[1]{\textcolor{cyan}{RC: #1}}
\newcommand{\toref}{{\color{red}[REF]}}

\bigskip
\begin{center}
{\Large
\noindent 
Limited knowledge exchange across biological subfields in agent-based modelling: 
an LLM-assisted bibliometric analysis}
\end{center}
\begin{flushleft}
\bigskip 
Kira Pugh\textsuperscript{1$\circledast$}, 
Rebecca M Crossley\textsuperscript{2}, 
Sara Hamis\textsuperscript{1}.

\bigskip
\textbf{1} Department of Information Technology, Uppsala University, Uppsala, Sweden. 
\\
\textbf{2} {Mathematical Institute, University of Oxford, Oxford, United Kingdom.}
\\

\end{flushleft}

\smallskip
\noindent{\bf Author ORCiDs}: 
\smallskip
\textbf{KP:} 0000-0001-5588-7014; 
\textbf{RC:} 0000-0001-7342-0207;
\textbf{SH:} 0000-0002-1105-8078.

\bigskip
\noindent{\bf $\circledast$ Corresponding author}: kira.pugh@it.uu.se

\thispagestyle{empty} 
\vspace{5cm}
\section*{Abstract}
Agent-based models (ABMs) are widely used across the biological sciences.
While applications differ between biological subfields, the mathematical and statistical methods underlying ABMs largely do not, raising the question of how much knowledge is exchanged between subfields.
We address this question by developing a bibliometric pipeline that combines a database keyword search with coarse- and fine-grained classification by a large language model (LLM).
We apply the pipeline to the Web of Science Core Collection, classifying 22,936 ABM articles into ecology, epidemiology, oncology, other-biology, or not-biology. 
Using citation patterns and author overlap as proxies for knowledge exchange, we find that 83\%, 81\%, and 69\% of citations from ecology, epidemiology, and oncology articles, respectively, go to articles in the same subfield. 
Furthermore, 91\% of authors with at least two articles across these subfields publish exclusively within one subfield. 
Both proportions far exceed those expected under random assignment of subfields (permutation tests, $p \leq 0.001$).
Terminology also differs, with ``individual-based model'' used mainly in ecology and ``agent-based model'' in epidemiology, oncology, and outside biology.
Taken together, our results suggest limited knowledge exchange between biological subfields in ABM research, leading us to discuss opportunities for increased exchange through conferences, journals, and training.

\bigskip
\noindent{\bf Keywords}:  
agent-based modelling, scientometrics, bibliometric analysis, mathematical biology
\newpage 
\clearpage
\pagenumbering{arabic}

\vspace*{0.25cm}

\section{Introduction}
Agent-based models (ABMs), also known as individual-based models (IBMs), are a mathematical and computational framework used to simulate autonomous agents that interact with each other and with their environment. 
Each agent follows deterministic or stochastic rules and has its own attributes (e.g., type, position, and age), allowing for heterogeneous populations of agents~\cite{crooks2011introduction}.
ABMs can therefore be used to study how population-level dynamics emerge from individual-level behaviours, even in systems that are difficult to study analytically~\cite{bonabeau2002agent,macal2009agent,macal2010tutorial}.
The flexibility of ABMs has made them a widely used tool across the biological sciences, including in ecology~\cite{deangelis2005individual,zhang2020overview}, epidemiology~\cite{tracy2018agent,auchincloss2008new}, and oncology~\cite{metzcar2019review,norton2019multiscale}, where agents may represent, e.g., cells, plants, animals, or humans~\cite{macal2010tutorial}.

Modern ABMs have their origins in rule-based systems such as cellular automata (CA), which were developed in the 1940s by pioneering mathematicians including John von Neumann and Stanisław Ulam, partly to investigate the principles underlying biological self-reproduction~\cite{ilachinski2001cellular,sarkar2000brief,vonNeumann1966}. 
CA reached a wider audience in 1970, when Martin Gardner popularised John Conway's {\it Game of Life}, in which simple rules governing the birth, death, and survival of organisms on a grid generate complex patterns \cite{payette2011agent,gardner1970mathematical}.
Around the same time, Thomas Schelling brought rule-based modelling into the social sciences with his model of residential segregation~\cite{schelling1969,schelling1971dynamic}.
Stephen Wolfram later demonstrated that even simple CA can generate highly complex behaviour~\cite{wolfram1982,wolfram1984universality}, and with the spread of personal computers, ABMs became common across many disciplines in the 1990s~\cite{gilbert2002platforms}.

Although the research questions differ between scientific disciplines, the mathematical and statistical methods underlying ABMs are largely shared, such that many methodological advances developed in one discipline should be transferable to others.
However, communities of ABM researchers have to a large extent emerged in parallel: individual-based modelling in ecology developed independently of agent-based modelling in the social and computational sciences~\cite{vincenot2018new}, and the two terms remain associated with different research communities today~\cite{grimm2005,an2021challenges}.
This parallel development raises the question of how much knowledge is exchanged between ABM communities in different disciplines or, as is the focus of this study, between biological subfields.

Using citation patterns and author overlap across subfields as example proxies, knowledge exchange can be studied through bibliometric analysis, which uses statistical methods to examine patterns in publication data \cite{ding2025insights,donthu2021conduct}. 
These data can be obtained from citation databases such as the Web of Science~\cite{webofscience}, Scopus~\cite{scopus}, and OpenAlex~\cite{OpenAlex}.
A search of the Web of Science Core Collection (WoSCC)~\cite{webofsciencecorecollection} identified 47 bibliometric studies of ABMs (Supplementary Material, SM1), including studies of ABMs in epidemiology~\cite{tang2022agent}, transportation~\cite{domenteanu2023data}, and finance~\cite{trinidad2022bibliometric}. 
Moreover, literature reviews of biology-focused ABMs that include bibliometric features have been conducted within individual subfields, such as ecology~\cite{grimm1999ten}, epidemiology~\cite{willem2017lessons}, and oncology~\cite{stephan2024agent}.
In contrast, Vincenot~\cite{vincenot2018new} analysed citation patterns of ABM research across disciplines using journal-level subject categories that group biological research into pre-defined areas such as ``agricultural and biological sciences'' and ``medicine''.
Separating articles into finer biological subfields is challenging using traditional bibliometric approaches, as keyword searches rely on explicit terminology~\cite{xu2024automating} and database subject categories can be too broad~\cite{waltman2012new}. 
However, recent advances in large language models (LLMs) offer an alternative approach, allowing users to classify articles based on the semantic content of their text~\cite{santos2024large,sun2023text,kostina2025large}.

In this work, we develop an LLM-assisted bibliometric pipeline to study ABM research across biological subfields.
We focus on ecology, epidemiology, and oncology: three subfields in which ABMs are a well-established modelling framework~\cite{white2021charting}. 
We compare terminology across subfields and use cross-subfield citation patterns and author overlap to assess knowledge exchange.

\section{Methods}
\label{sec:methods}

\subsection{Data collection and cleaning}
\label{sec:data}
\begin{figure}[b!]
    \centering
    \includegraphics[width=1\linewidth]{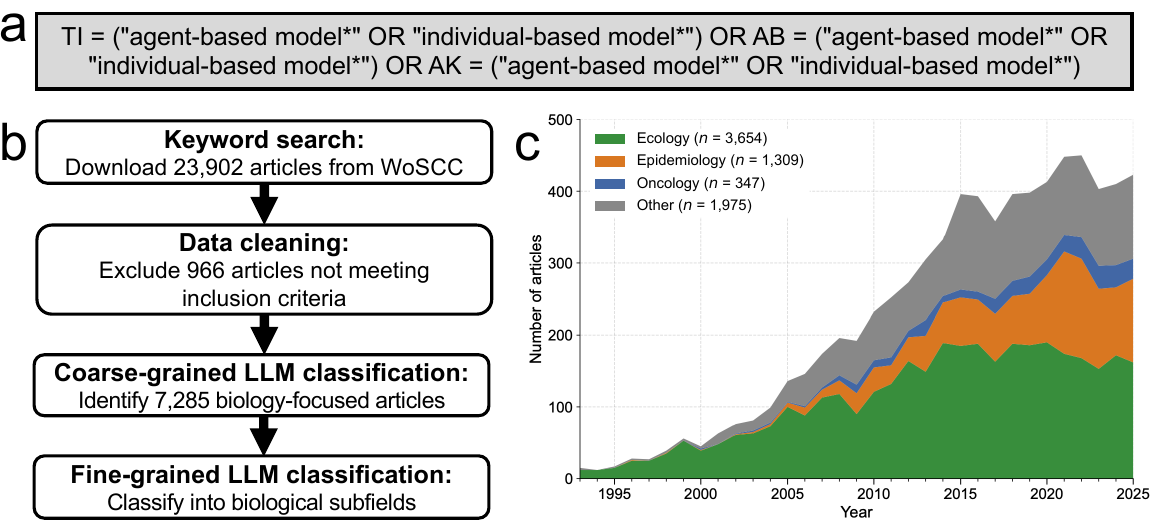}
    \caption{
    \textbf{Article selection and classification.}
    (a) Search query used to retrieve documents from the WoSCC. 
    TI, AB, and AK denote the article title, abstract, and author keywords, respectively. 
    The asterisk denotes a wildcard that matches any word ending. 
    (b) Data collection, cleaning, and classification workflow. 
    (c) Number of articles per year in each biological subfield, shown as a stacked area chart.
    $n$ denotes the total number of articles in each subfield. 
    }
    \label{fig:flowchart}
\end{figure}

We used the Web of Science (WoS) application programming interface (API)~\cite{clarivate_wos_api} to retrieve metadata for all documents in the WoSCC containing ``agent-based model'' or ``individual-based model'' in the title, abstract, or author keywords (Figure~\ref{fig:flowchart}a).
For each of the 23,902 documents retrieved on 9 January 2026, we obtained the following metadata: title, abstract, author keywords, publication year, author ORCID iDs, and references. 
To ensure sufficient data per year, we excluded the 12 documents published before 1993, the first year with more than 10 publications. 
We also removed 755 documents with missing titles or abstracts and 199 non-English documents (Figure~\ref{fig:flowchart}b), resulting in a dataset of 22,936 ABM articles published between 1993 and 2025.

\subsection{Article classification}
\label{sec:classification}
To classify the articles, we used the LLM Qwen3:8b~\cite{bai2023qwen}, 4-bit quantised and run with Ollama (v0.40.1), with temperature 0. 
Two zero-shot prompts~\cite{brown2020language,wang2023large}, written in American English~\cite{nayeem2026english}, were used. 
Prompt 1 separated biology from not-biology articles (Section~\ref{sec:coarse}, Appendix~\ref{app:prompt1}), and Prompt 2 categorised the biology articles into four prescribed biological subfields: epidemiology, oncology, ecology, or other-biology (Section~\ref{sec:fine}, Appendix~\ref{app:prompt2}). 
Each prompt was refined iteratively~\cite{schulhoff2024prompt,liu2023pre,mu2024navigating} on a human-annotated development set sampled from the dataset it was applied to, and was then evaluated on separate human-annotated data (Section~\ref{sec:eval}).

\subsubsection{Coarse-grained classification: biology versus not-biology}
\label{sec:coarse}
Prompt 1 (Appendix~\ref{app:prompt1}) instructed the LLM to classify each article abstract as biology or not-biology based on its primary scientific focus. 
Accordingly, biology was defined as research on living organisms or biological processes, in line with the common definition of biology as the scientific study of life~\cite{taylor2021campbell}, and not-biology as research on physical systems, engineering, modelling, social sciences, or environmental processes without biological mechanisms.
The prompt was refined on a development set of 100 randomly sampled articles (without stratification), which one author annotated as 36 biology and 64 not-biology articles. 
Refinement stopped when the agreement between the human annotations and the LLM classifications reached an accuracy of 94\%. 
The WoS unique identifiers, DOIs (where available), and annotations are listed in Supplementary Table SM1. 
Applied to the full dataset (including the development set), the prompt identified 7,285 biology and 15,651 not-biology articles.

\subsubsection{Fine-grained classification: biological subfields}
\label{sec:fine}

Prompt 2 (Appendix~\ref{app:prompt2}) instructed the LLM to categorise each biology-classified article into exactly one biological subfield based on the primary focus of its abstract.
We specified three main subfields: (1) oncology, if the article focuses on cancer, (2) epidemiology, if it focuses on disease spread or population-level health outcomes in humans or animals, and (3) ecology, if it focuses on interactions among living organisms and their environment; (4) other-biology is a residual category for articles that do not fit these. 
The prompt applied these in the listed order, prioritising the three main subfields in an effort to minimise the number of incorrectly classified articles in them.
The prompt was refined on a development set of 100 randomly sampled biology articles (without stratification), which one author annotated as 54 ecology, 20 epidemiology, 5 oncology, and 21 other-biology articles.
Refinement stopped once agreement with the human annotations plateaued at an overall accuracy of approximately 90\% (and a precision of 95\% across ecology, epidemiology, and oncology).
The WoS unique identifiers, DOIs (where available), and annotations are listed in Supplementary Table SM2.
Applied to all 7,285 biology articles (including the development set), the prompt identified 3,654 ecology, 1,309 epidemiology, 347 oncology, and 1,975 other-biology articles.
Figure~\ref{fig:flowchart}c shows that ecology was the subfield with the largest number of ABM-related articles published throughout our dataset. 
Before the year 2000, over 80\% of yearly ABM-related biology articles were ecology-based, with this proportion remaining above 50\% until 2008, and never falling below 37\% throughout the study period.

\subsubsection{LLM evaluation}
\label{sec:eval}
To evaluate Prompt 1, the three authors collectively annotated a total of 150 abstracts predicted to be biology and 150 predicted to be not-biology. 
The abstracts were randomly sampled within each predicted category, and the annotators were blind to the LLM predictions. 
Each author independently annotated 100 abstracts from each category, with 50 shared between each pair of authors.
Disagreements were first reconsidered by the two authors involved after rereading the abstract, and any remaining ties were broken by the third author.
The author pairs initially agreed on 96\% of the abstracts. 
Of the 12 disagreements, 11 were resolved by rereading and 1 by the third author. 
All LLM classifications and human annotations (initial, after rereading, and tie-break) are listed in Supplementary Table SM3.
As abstracts were sampled by predicted category, we evaluated the prompt using two metrics: precision (the percentage of articles predicted as biology that were annotated as biology) and the negative predictive value (NPV; the percentage of articles predicted as not-biology that were annotated as not-biology). 
The confusion matrix (Table~\ref{tab:confusion_matrix_prompt1}) gives a precision of 95\% and an NPV of 89\%. 
We considered precision more important, since a false positive contaminates the subfield analysis, whereas a false negative mainly reduces the sample size.
This comes at a cost in coverage: based on the NPV, we estimated that approximately 11\% (about 1,700) of the 15,651 articles classified as not-biology are in fact biology.

\begin{table}[h]
\centering
\begin{tabularx}{0.99\textwidth}{|c|>{\raggedright\arraybackslash\hsize=0.58\hsize}X|*{2}{>{\centering\arraybackslash\hsize=1.21\hsize}X|}}
\cline{3-4}
\multicolumn{2}{c|}{}
& \multicolumn{2}{c|}{\textbf{Annotated}} \\
\cline{3-4}
\multicolumn{2}{c|}{}
& \textbf{Biology}
& \textbf{Not-biology} \\
\hline
\multirow[c]{2}{*}{\hspace{-2pt}\raisebox{-7pt}[0pt][0pt]{\rotatebox[origin=c]{90}{\textbf{Predicted}}}}
& \textbf{Biology}
& \rule[-10pt]{0pt}{28pt}142 (94.67\%)
& 8 (5.33\%) \\
\cline{2-4}
& \textbf{Not-biology}
& \rule[-10pt]{0pt}{28pt}16 (10.67\%)
& 134 (89.33\%) \\
\hline
\end{tabularx}
\caption{\textbf{Confusion matrix for Prompt 1.}
Percentages are calculated relative to the total number of articles in each predicted category ($n=150$).
The top-left percentage is the precision and the bottom-right is the negative predictive value.}
\label{tab:confusion_matrix_prompt1}
\end{table}

\newpage

Prompt 2 was evaluated similarly.
Each author independently annotated 50 abstracts from each predicted subfield, with 25 shared between each pair of authors, totalling 75 distinct abstracts per subfield.
The author pairs initially agreed on 86\% of the abstracts. Of the 43 disagreements, 24 were resolved by rereading and 19 by the third author, 18 of which involved other-biology.
All LLM classifications and human annotations are listed in Supplementary Table SM4, and the evaluation results are summarised in the confusion matrix (Table~\ref{tab:confusion_matrix_prompt2}).
We evaluated Prompt 2 using the precision of each predicted subfield, which is high for ecology (93\%), epidemiology (95\%), and oncology (100\%), and lower for other-biology (72\%), where 21 of the 75 abstracts were annotated as ecology (13), epidemiology (7), or oncology (1), consistent with its role as the residual category.
Although the stringent, priority-based assignment placed some main-subfield articles in other-biology, the high precision meant that each main subfield contained few articles that belong elsewhere, yielding reliable datasets for investigating cross-subfield knowledge exchange.

\begin{table}[ht]
\centering
\begin{tabularx}{0.99\textwidth}{|c|>{\raggedright\arraybackslash}X|*{4}{>{\centering\arraybackslash}X|}}
\cline{3-6}
\multicolumn{2}{c|}{}
& \multicolumn{4}{c|}{\textbf{Annotated}} \\
\cline{3-6}
\multicolumn{2}{c|}{}
& \textbf{Ecology}
& \mbox{\textbf{Epidemiology}}
& \textbf{Oncology}
& \textbf{Other} \\
\hline
\multirow{4}{*}{\rotatebox[origin=c]{90}{\textbf{Predicted}}}
& \textbf{Ecology}
& 70 (93.33\%)
& 1 (1.33\%)
& 0 (0.00\%)
& 4 (5.33\%) \\
\cline{2-6}
& \mbox{\textbf{Epidemiology}}
& 0 (0.00\%)
& 71 (94.67\%)
& 0 (0.00\%)
& 4 (5.33\%) \\
\cline{2-6}
& \textbf{Oncology}
& 0 (0.00\%)
& 0 (0.00\%)
& 75 (100.00\%)
& 0 (0.00\%) \\
\cline{2-6}
& \textbf{Other}
& 13 (17.33\%)
& 7 (9.33\%)
& 1 (1.33\%)
& 54 (72.00\%) \\
\hline
\end{tabularx}

\caption{\textbf{Multiclass confusion matrix for Prompt 2.}
Percentages are calculated relative to the total number of articles in each predicted subfield ($n=75$).
The diagonal percentages are the precision of each predicted subfield.}
\label{tab:confusion_matrix_prompt2}
\end{table}

\subsection{Bibliometric analyses}
\label{sec:analyses}

\subsubsection{Terminology}
\label{sec:terminology}

To compare terminology across subfields, we used case- and hyphen-insensitive matching to classify each article according to the search terms in its title, abstract, or author keywords: only ``agent-based model*'', only ``individual-based model*'', or both.
Here, * denotes a wildcard matching any word ending, such as ``models'' or ``modelling''. 
We then calculated the number of articles in each category per year for each of the four biological subfields, and for not-biology articles as a reference.

\subsubsection{Citation patterns}
\label{sec:citations}
To study knowledge exchange through citations, we analysed citations between articles within the biology dataset, excluding any articles outside it. 
We included all four fine-grained categories, as excluding the residual other-biology category would inflate within-subfield citation proportions.
For each subfield pair (including self-pairs), we calculated the number of citations given by articles in the first subfield and received by articles in the second.

To test whether within-subfield citation proportions were higher than expected from subfield sizes alone, that is, under random assignment of subfields, we performed a permutation test~\cite{nichols2002nonparametric}. 
In each permutation, subfield labels were randomly shuffled among articles published in the same year, while citation links between articles remained unchanged.
This preserved annual subfield sizes and the citation network, but removed the association between individual articles and their subfields.
Within-subfield citation proportions were recalculated for $B = 1000$ permutations, giving a null distribution for each subfield.
For the one-sided alternative hypothesis that the observed within-subfield citation proportion exceeds that under random assignment of subfield labels, the $p$-value was approximated by the Monte Carlo $p$-value $p = (b+1)/(B+1)$~\cite{phipson2010permutation}. 
Here, $b$ is the number of permutations in which the within-subfield citation proportion was at least as high as the observed proportion, and adding one to both the numerator and the denominator, which counts the observed data as one permutation, ensures a valid $p$-value that is never zero~\cite{north2002note,phipson2010permutation}.

\subsubsection{Author overlap}
\label{sec:authors}
To study knowledge exchange through authors who publish in multiple subfields, we identified authors by their ORCID iDs.
We restricted this analysis to articles in the biology dataset published from 2010 onwards, when annual ORCID coverage was at least 50\% in all subfields (Supplementary Material SM3.1).
%
Unlike in the citation analysis, we also excluded the residual other-biology category, which is heterogeneous by design and contains some misclassified articles from the three main subfields (Section~\ref{sec:eval}).
Thus, author overlap with the other-biology category would overstate knowledge exchange between research communities.
To test the robustness of our results to this exclusion, we repeated the analysis with other-biology included (Supplementary Material SM3.2). 
Furthermore, only authors with at least two articles across the three main subfields (ecology, epidemiology, and oncology) were included, since authors with a single article cannot publish in more than one subfield.
For each subfield, we calculated author overlap as the percentage of its authors who also published in at least one of the other two subfields.

To test whether author overlap was lower than expected from authors' publication counts alone, that is, under random assignment of subfields, we used a permutation test analogous to that for citations (Section~\ref{sec:citations}).
In each permutation, subfield labels were randomly shuffled among articles published in the same year, while the links between authors and articles remained unchanged.
This preserved annual subfield sizes and the number and timing of each author's articles, but removed the association between individual articles and their subfields.
Author overlap was recalculated for $B = 1000$ permutations, giving a null distribution for each subfield. 
For the one-sided alternative hypothesis that the observed author overlap is lower than under random assignment of subfield labels, the $p$-value was approximated by the Monte Carlo $p$-value $p = (b+1)/(B+1)$~\cite{phipson2010permutation}, where $b$ is the number of permutations in which author overlap was at least as low as observed.

\section{Results}
\label{sec:results}
\enlargethispage{2\baselineskip}
\subsection{ABM or IBM? Terminology differs between subfields}

\begin{figure}[h!]
    \centering
    \includegraphics[width=1\linewidth]{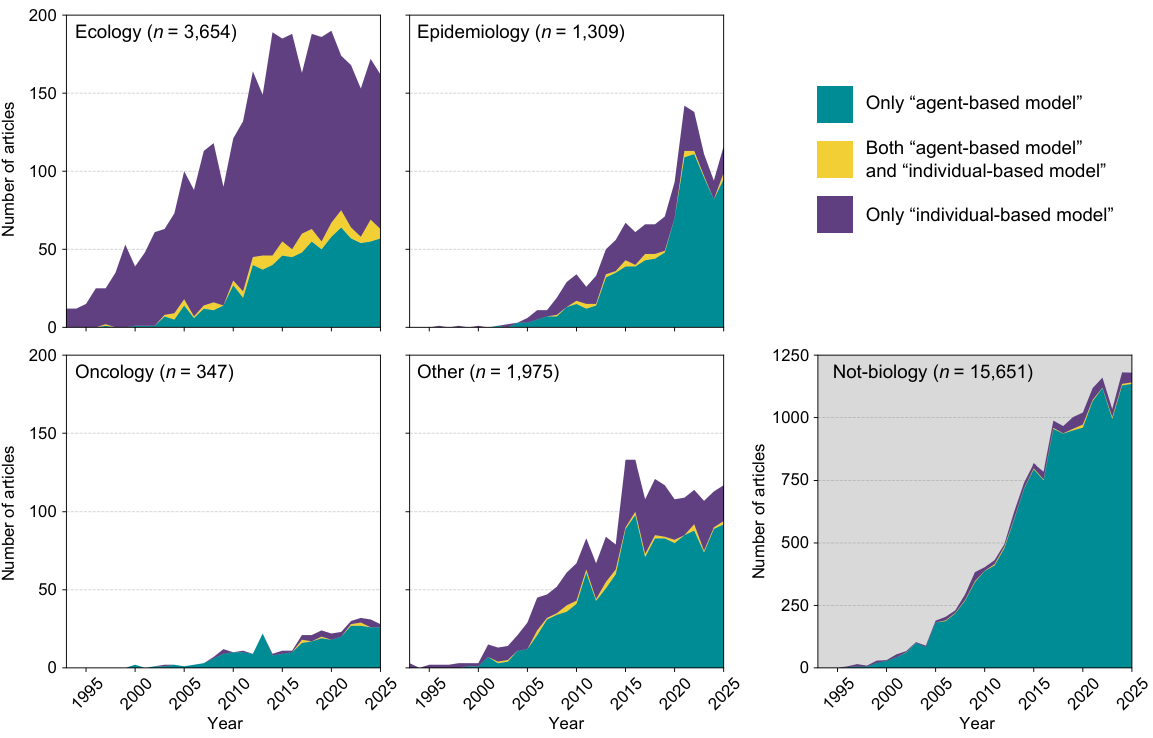}
    \caption{
    \textbf{Terminology used across biological subfields and not-biology articles.} 
    Annual number of articles containing ``agent-based model" only, ``individual-based model" only, or both terms together, in the title, abstract, or author keywords, across biological subfields and not-biology. 
    $n$ denotes the total number of articles in each category. 
    }
    \label{fig:ABM_versus_IBM}
\end{figure}

\label{sec:res_terminology}
Figure~\ref{fig:ABM_versus_IBM} shows the annual number of articles using ``agent-based model'', ``individual-based model'', or both terms, for each biological subfield and for not-biology articles (raw counts in Supplementary Table SM5).
In not-biology articles, ``agent-based model'' dominated throughout the study period and ``individual-based model'' was rarely used.
Within biology, ``individual-based model'' was the most frequently used terminology in ecology-classified articles between 1993 and 2025. 
However, ``agent-based model'' dominated in both epidemiology and oncology.
In absolute terms, the use of ``agent-based model'' increased in all biological subfields across the study period, whereas the use of ``individual-based model'' grew markedly only in ecology.
In relative terms, epidemiology shifted from preferring one term to the other: ``individual-based model'' was more frequently used until around 2010, after which ``agent-based model'' became predominant, reaching 87\% of articles (82 of 94) in 2024.
Articles using both terms together were rare (ranging from 0 to 12\%) in all subfields and years.

\subsection{Subfields mainly cite themselves}
\begin{figure}[h!]
    \centering
    \includegraphics[width=1\linewidth]{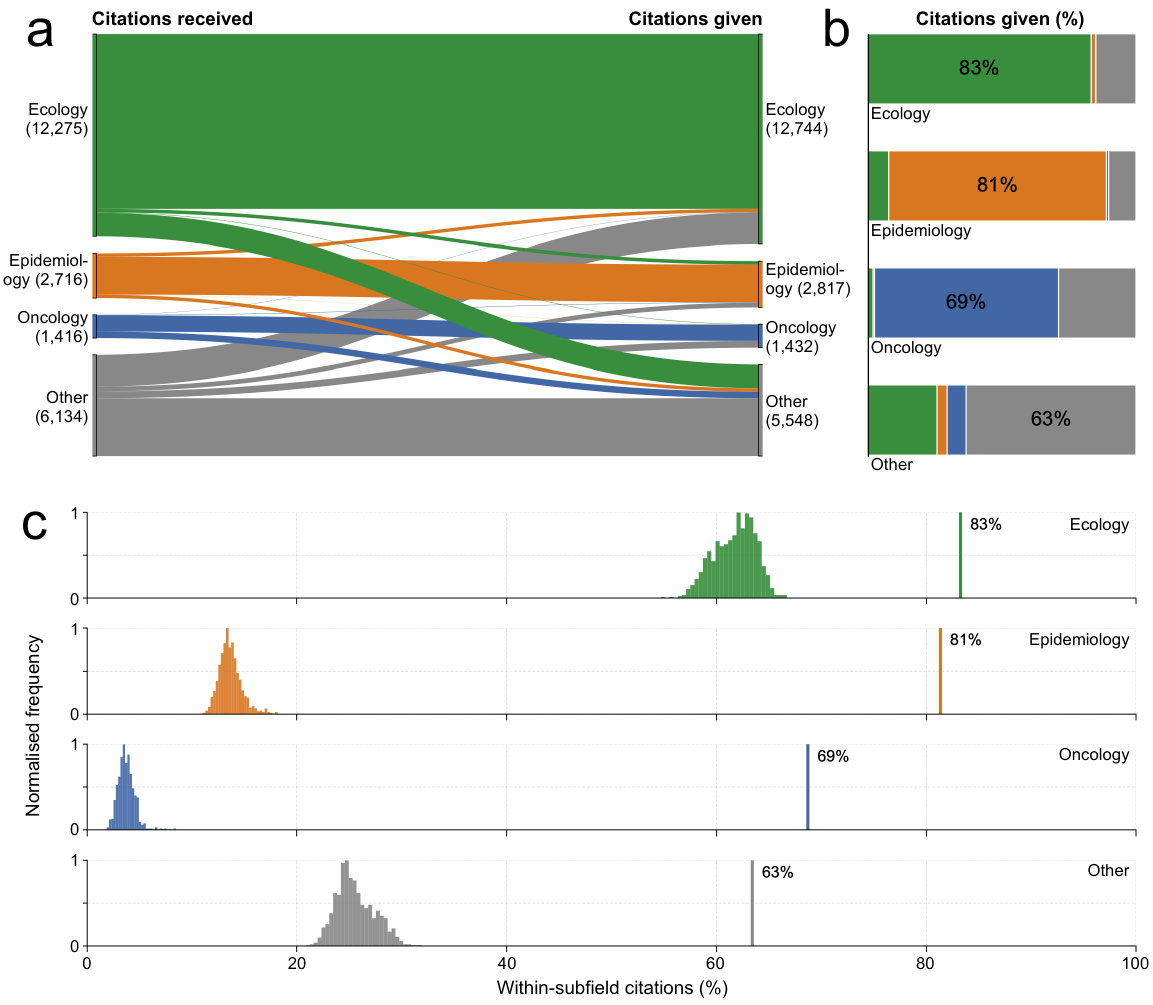}
    \caption{\textbf{Citation flows within and between biological subfields.}
    Subfields are ecology, epidemiology, oncology, and the residual other-biology category, and only citations between articles in the biology dataset are included.
   (a) Citation flows between subfields.
    Node heights are proportional to the total number of citations received (left) or given (right) by each subfield, and ribbon heights are proportional to the number of citations between subfields.
    (b) Proportion of citations given to each cited subfield, with colours as in (a).
    (c) Null distributions of within-subfield citation proportions from 1000 permutations, in which subfield labels were shuffled among articles within each publication year.
    Vertical lines mark the observed values.
    }
    \label{fig:cross_talk}
\end{figure}

\label{sec:res_citations}
Figure~\ref{fig:cross_talk} shows citation flows between biological subfields.
In absolute terms, ecology gave and received the most citations, whereas oncology exchanged few, consistent with their respective sizes (Figure~\ref{fig:cross_talk}a). 
Relative to subfield size, citations were concentrated within subfields: 83\%, 81\%, 69\%, and 63\% of citations given by ecology, epidemiology, oncology, and other-biology, respectively, went to articles in the same subfield (Figure~\ref{fig:cross_talk}b).
Where cross-subfield citations occurred, exchange was largely reciprocal (Supplementary Material SM2).
For all four subfields, the observed within-subfield citation proportion lay far above the null distribution, whose central 95\% ranges were [57.77\%, 64.90\%], [11.88\%, 16.42\%], [2.34\%, 5.45\%], and [22.50\%, 29.36\%] for ecology, epidemiology, oncology, and other-biology, respectively (Figure~\ref{fig:cross_talk}c).
No permutation reached the observed values ($p \leq 0.001$ for all subfields).
The concentration of citations within subfields therefore cannot be explained by differences in subfield size and publication year alone.

\subsection{Authors mainly publish within one subfield}
\label{sec:res_authors}
\begin{figure}[H]
\centering
\includegraphics[width=\linewidth]{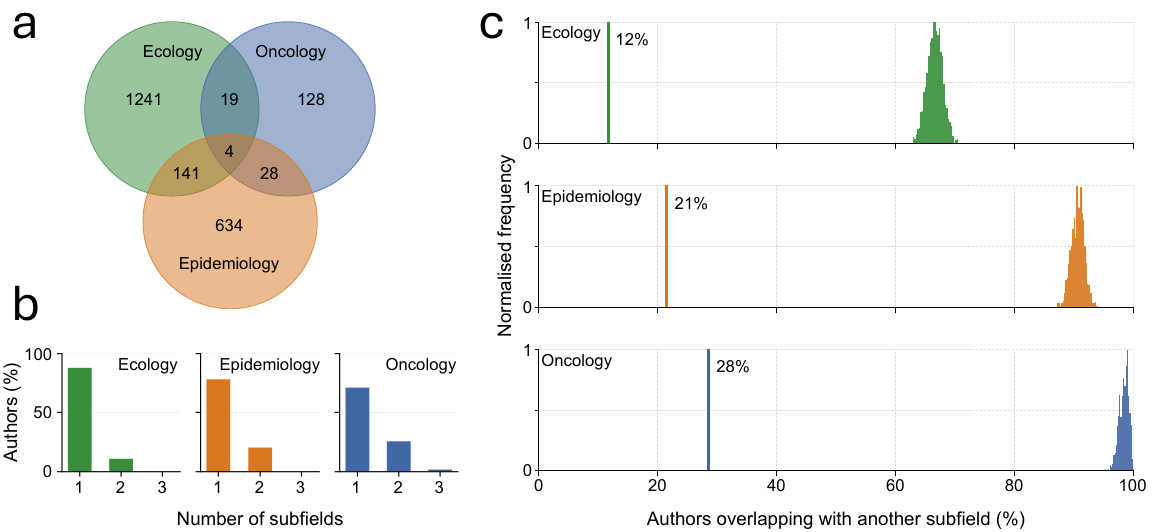}
\caption{\textbf{Authors publishing within and across biological subfields.}
Subfields are ecology, epidemiology, and oncology, and only authors with an ORCID iD and at least two articles across these subfields published between 2010 and 2025 are included.
(a) Number of unique authors publishing in each combination of subfields.
(b) For authors who published in ecology, epidemiology, and oncology (bars, left to right), the proportion who published in one, two, or three subfields. Colours as in (a). 
(c) Null distributions of author overlap from 1000 permutations, in which subfield labels were shuffled among the included articles within each publication year.
Vertical lines mark the observed values.
}
\label{fig:orcid_venn}
\end{figure}

Figure~\ref{fig:orcid_venn} shows author overlap between ecology, epidemiology, and oncology for the 2,195 authors with at least two articles across these subfields from 2010 onwards.
Of these, 91\% published in only one subfield (Figure~\ref{fig:orcid_venn}a,b).
The percentage of authors who also published in at least one of the other two subfields was 12\%, 21\%, and 28\% for ecology, epidemiology, and oncology, respectively (Figure~\ref{fig:orcid_venn}c).
Relative overlap was lowest for ecology, the largest subfield, and highest for oncology, the smallest.

For all three subfields, observed author overlap lay far below the null distribution, whose central 95\% ranges were [64.02\%, 69.22\%], [88.52\%, 92.72\%], and [96.67\%, 99.57\%], respectively (Figure~\ref{fig:orcid_venn}c). 
In none of the 1000 permutations was author overlap as low as observed, giving $p \leq 0.001$ for all three subfields.
When the analysis was repeated to include the residual other-biology category, overlap increased to 34\%, 36\%, and 66\% for ecology, epidemiology, and oncology, respectively, mainly through authors who also published in other-biology (Supplementary Material SM3.2).
Nevertheless, overlap remained far below the null distribution for all subfields ($p \leq 0.001$).
Within our dataset, authors therefore published across subfields considerably less often than they would have if their articles had been randomly assigned to subfields.

\section{Discussion}
In this study, we developed an LLM-assisted bibliometric pipeline to classify ABM research into biological subfields.
We compared terminology across these subfields and used citation patterns and author overlap as proxies for knowledge exchange between them.
Three main findings stand out.
Firstly, terminology differed between subfields: ``individual-based model'' was used mainly in ecology, whereas ``agent-based model'' was more common in epidemiology and oncology (Section~\ref{sec:res_terminology}).
Secondly, between 69\% and 83\% of citations given by ecology, epidemiology, and oncology articles went to articles in the same subfield (Section~\ref{sec:res_citations}).
Thirdly, 91\% of authors with at least two articles in these subfields published in only one of them (Section~\ref{sec:res_authors}).
In both the citation and author analyses, the separation between subfields was greater than under random assignment of subfields.
Together, these patterns indicate that biological ABM research is organised into largely distinct subfield communities.

The observed separation between subfields is particularly notable because ABMs share core mathematical, statistical, and computational foundations across applications~\cite{macal2010tutorial,grimm2005}, so methods developed in one subfield could reasonably be expected to be useful in others.
Although our bibliometric analyses cannot explain why the separation exists, the differences in terminology indicate distinct modelling traditions in different subfields, which may contribute to it.
Terminology differences can make relevant literature harder to find across subfields~\cite{Bracken2006} and may thereby reinforce the separation between communities.
Separation between research communities has also been observed beyond ABMs, with fields often becoming increasingly isolated as they mature~\cite{petterson2021measuring,singh2022quantifying}, although our previous bibliometric study suggests that mathematical oncology has become more interconnected over time~\cite{pugh2025bibliometric}. 

The results presented in this study should be interpreted in light of several design choices and their associated limitations.
The keyword search used for article retrieval relied on the terms ``agent-based model'' and ``individual-based model'', so relevant articles using other terminology, such as early rule-based models~\cite{gardner1970mathematical,schelling1971dynamic}, were not included.
Keyword bias is a well-documented limitation of bibliometric studies, affecting which articles are included in the dataset~\cite{ozturk2024design,grames2019automated,Chen2016}, and the dataset itself was limited to documents indexed in the WoSCC~\cite{webofsciencecorecollection}.
Other bibliographic databases, such as Scopus~\cite{scopus} and OpenAlex~\cite{OpenAlex}, differ in coverage, particularly of conference proceedings and non-English-language journals~\cite{Visser2021}.
Similarly, the author analysis was limited to researchers with associated ORCID iDs.

The article classification also involved several simplifications.
Articles were assigned to exactly one subfield, with overlaps resolved by a fixed priority order (oncology, epidemiology, ecology) chosen in increasing order of subfield size, as estimated from keyword frequencies (Appendix~\ref{app:prompt2}).
Articles spanning several subfields, such as models of cancer epidemiology, were therefore counted in only one, so knowledge exchange within individual articles was not captured.
Classification was based only on abstracts, which generally capture an article's main focus but may omit details that would clarify its biological application.
Moreover, the decision to prioritise precision in the three main subfields over complete coverage came at the cost of lower precision for the residual other-biology category (72\%), which therefore contained articles that human annotators assigned to the main subfields.
Since any article misclassified in this way turns its within-subfield citations into cross-subfield citations, and may inflate author overlap in the analysis including other-biology (Supplementary Material SM3.2), our findings of limited knowledge exchange can be interpreted as conservative.
However, random assignment of subfields provides only a minimal baseline, and without comparison to other modelling approaches or fields, we cannot assess whether the observed separation in biological ABM research is unusually strong.

\enlargethispage{2\baselineskip}
Furthermore, using citations and author overlap within our dataset as proxies for knowledge exchange is itself a simplification.
Our analysis did not distinguish whether a citation built on, critiqued, or only briefly mentioned prior work, and citations to articles outside the dataset were not counted.
Bibliometric approaches that could investigate knowledge exchange in more depth include bibliographic coupling~\cite{Kessler1963,Boyack2010}, citation context analysis~\cite{bornmann2008citation,Tahamtan2019}, textual similarity of abstracts~\cite{Boyack2011}, models of topic diffusion between disciplines~\cite{kiss2010can}, and analyses of software mentions in articles~\cite{Howison2016}.
As a final limitation, our analysis identifies limited knowledge exchange but does not measure whether this has affected research quality or impact, which for interdisciplinary research can take longer to emerge~\cite{zhang2024delayed}.

\newpage
Future work could address the discussed limitations, for example by combining several bibliographic databases or by integrating multiple bibliometric approaches~\cite{Lim2024}.
Moreover, studying how citations and author overlap change over time could test whether knowledge exchange between subfields is increasing or decreasing.
Journal- or conference-level analyses could furthermore identify where cross-subfield communication occurs and, equally importantly, where it does not.
Comparing ABMs with other modelling approaches, such as differential equation models, could also show whether the observed separation is specific to ABMs or characteristic of quantitative biology in general. 
The funnel-shaped article classification pipeline developed in this study could be applied to other research methods and disciplines, especially since the computationally expensive LLM steps are applied only to progressively smaller subsets of articles, allowing it to scale to large datasets. 
Any such reuse should include transparent reporting of methodological choices and evaluation against human annotations, in line with recommended practices for LLM-based annotation~\cite{Trnberg2024}. 
Beyond bibliometrics and the quantitative proxies used for knowledge exchange in this work, qualitative approaches such as surveys~\cite{Rijnsoever2011} and interviews~\cite{MacLeod2018} could clarify how researchers discover work across subfields and which barriers they face. 
Such approaches could also help identify examples of knowledge exchange that have proved useful in practice.

While our work identifies, and largely focuses on, limited knowledge exchange between subfields, we stress that subfield-focused research has many benefits, such as deep domain expertise, shared vocabulary, and specialist peer review~\cite{jacobs2013}.
Interdisciplinary research, on the other hand, has been noted to carry costs, including lower funding success~\cite{Bromham2016} and fewer publications, albeit with higher visibility~\cite{Leahey2017}.
Several studies have investigated whether interdisciplinary research receives more citations or produces better science, often with contradictory results~\cite{levitt2008multidisciplinary,lariviere2015long,okamura2019interdisciplinarity,wang2015interdisciplinarity}.
Specialised research communities also benefit from communication in dedicated modelling journals: mathematical ecologists may publish in \textit{Ecological Modelling} and \textit{Theoretical Ecology}, and mathematical epidemiologists in \textit{Epidemics} and \textit{Infectious Disease Modelling}.
Mathematical oncologists, however, mainly publish in broader journals, as no dedicated mathematical oncology journal has yet become established.
Specialisation is also reflected in ABM software.
Although general-purpose platforms such as NetLogo~\cite{wilensky1999netlogo}, Mesa~\cite{terhoeven2025mesa}, and Repast Simphony~\cite{north2013complex} can be applied across biological subfields, more specialised platforms have been developed in parallel for specific applications. 
For example, HexSim~\cite{schumaker2018hexsim} and RangeShifter~\cite{bocedi2014rangeshifter} are widely used in ecology, Covasim~\cite{kerr2021covasim} and Epidemiological Modeling software (EMOD)~\cite{bershteyn2018implementation} in epidemiology, and PhysiCell~\cite{ghaffarizadeh2018physicell}, CompuCell3D~\cite{swat2012compucell3d}, Chaste~\cite{mirams2013chaste}, and the Hybrid Automata Library (HAL)~\cite{bravo2020hybrid} in oncology.

We argue that knowledge exchange between subfields can be valuable, and that there are many opportunities to increase such exchange while retaining subfield-specific expertise.
Interdisciplinary workshops, methods-focused training, and special issues in scientific journals could make methods and terminology visible across subfields. 
Notably, the Ten Simple Rules collection in \textit{PLOS Computational Biology} offers practical guidance on interdisciplinary collaboration~\cite{knapp2015ten,carey2018ten,stawarczyk2023establishing,robinson2024ten}.
Furthermore, broad mathematical biology meetings, such as those organised by the \textit{Society for Mathematical Biology} and the \textit{SIAM Life Sciences} community, bring together researchers working on ecological, epidemiological, and oncological models while also providing subfield-focused symposia and subgroups activities. 
By connecting researchers across biological subfields, such initiatives can help turn shared methodological foundations into shared scientific progress. 

\subsection*{Data accessibility}

The code to reproduce the data cleaning, article classification, and results in this study is available on the public GitHub repository \url{https://github.com/KiraPugh/ABM_knowledge_exchange}. 
Data included in this study are derived from the WoS (Clarivate) and cannot be redistributed under its licence terms.
The search query and retrieval date are provided in Section \ref{sec:data}. The WoS unique identifiers and the DOIs (where available) of all articles making up the dataset are available in Supplementary Table 6.

\subsection*{Declaration of AI use}
The authors used AI to edit prose and code. All conceptual, methodological, theoretical, and empirical contributions are the authors’ own.

\subsection*{Conflict of interest declaration}
The authors declare no competing interests.

\subsection*{Funding}
KP was funded by the Wenner-Gren Stiftelserna/the Wenner-Gren Foundations through a grant to SH (WUP2024-0006).
RMC was supported by an Engineering and Physical Sciences Research Council grant (EP/Z534870/1).
SH was funded by Wenner-Gren Stiftelserna/the Wenner-Gren Foundations (WGF2022-0044) and the Kjell och M{\"a}rta Beijer Foundation.

\begin{spacing}{0.98}
\renewbibmacro{in:}{,}
\section*{References}
\vspace{-.2cm}
\AtNextBibliography{\small}
\printbibliography[heading=none]
\end{spacing}

\newpage
\appendix
\renewcommand{\thesection}{A\arabic{section}}
\section*{Appendix}

\section{Prompt 1}
\label{app:prompt1}

\refstepcounter{systemprompt}
\begin{tcolorbox}[colback=gray!10, colframe=black, title=Prompt \thesystemprompt\ (biology versus not-biology classification)]
\label{box:systemprompt1}
\begin{verbatim}
"""
You are an expert scientific document classifier. Determine whether a 
scientific abstract is primarily about **Biology** or **Not Biology**.

Definitions:
- Biology = Research on living organisms or biological processes.
- Not Biology = Research primarily on physical systems, engineering, 
modeling, social sciences, or environmental processes without biological 
mechanisms.

Instructions:
1. Only classify abstracts based on their primary scientific focus.
2. Output ONLY JSON in this format:

[
  {"id": 1, "Category": "Biology"},
  {"id": 2, "Category": "Not Biology"}
]

Do NOT add commentary or explanations.
"""
\end{verbatim}
\end{tcolorbox}

\section{Prompt 2}
\label{app:prompt2}

\refstepcounter{systemprompt}
\begin{tcolorbox}[colback=gray!10, colframe=black, title=Prompt 2 (biology subfield classification)]
\label{box:systemprompt2}
\begin{verbatim}
"""
You are an expert biological sciences classifier.

Classify each abstract into exactly ONE PRIMARY category.

You MUST determine the PRIMARY scientific focus of the study, not 
secondary mentions or background context.

--------------------------------------------------
DECISION ORDER (follow strictly):

1. oncology
   Use ONLY if cancer or oncology is the central research subject.
   Cancer must be explicitly studied (e.g. tumors, carcinogenesis, cancer 
   treatment, cancer-specific therapies).
   Do NOT use if cancer is only mentioned as motivation or comparison.

2. epidemiology
   Use if the PRIMARY focus is disease or health outcomes analyzed across 
   human or animal populations, i.e., epidemiology.
   Do NOT use if disease is a minor component of ecological or 
   environmental dynamics.

3. ecology
   Use if the study focuses on how living organisms interact with each 
   other and their environment, i.e., ecology.
   Includes wildlife disease ONLY when ecological dynamics (not public 
   health) are the primary outcome.

4. other-biology
   Use if the study does NOT clearly fit oncology, epidemiology, or ecology.
   Includes all other biology subfields, review papers, or methods papers 
   without a clear focus on health-related event in a population, 
   ecosystem, or cancer.

--------------------------------------------------

Rules:
- Choose the MOST SPECIFIC applicable category.
- Do NOT infer intent beyond what is stated.
- Never guess.
- Respond ONLY in valid JSON as a list of objects:
  [{"id": 1, "KA": "oncology"}, ...]
"""
\end{verbatim}
\end{tcolorbox}

\end{document}